\RequirePackage[mathlines]{lineno}
\documentclass[amsmath,amssymb,twocolumn,prd,floatfix,showpacs, nofootinbib]{revtex4-1}
\usepackage{tabularx}
\usepackage{bm, color}
\usepackage{overpic,subfigure}
\usepackage{multirow}
\usepackage{array}
\usepackage{dcolumn}
\usepackage[symbol]{footmisc}
\usepackage{epstopdf}
\usepackage{lineno}
\usepackage{hyperref}
\usepackage{amsmath}
\usepackage{graphicx}
\hypersetup{hypertex=true,
	        colorlinks = true,
            linkcolor = blue,
            urlcolor = blue,
            citecolor = blue}
\newcommand{\BESIIIorcid}[1]{\href{https://orcid.org/#1}{\hspace*{0.1em}\raisebox{-0.45ex}{\includegraphics[width=1em]{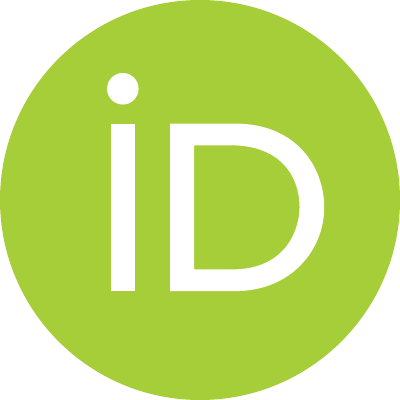}}}}

\begin{document}

\title{\boldmath Observation and branching fraction measurements of $\chi_{cJ}\to p \bar p K^0_S K^0_S$ }

\author{
M.~Ablikim$^{1}$\BESIIIorcid{0000-0002-3935-619X},
M.~N.~Achasov$^{4,c}$\BESIIIorcid{0000-0002-9400-8622},
P.~Adlarson$^{81}$\BESIIIorcid{0000-0001-6280-3851},
X.~C.~Ai$^{86}$\BESIIIorcid{0000-0003-3856-2415},
C.~S.~Akondi$^{31A,31B}$\BESIIIorcid{0000-0001-6303-5217},
R.~Aliberti$^{39}$\BESIIIorcid{0000-0003-3500-4012},
A.~Amoroso$^{80A,80C}$\BESIIIorcid{0000-0002-3095-8610},
Q.~An$^{77,64,\dagger}$,
Y.~H.~An$^{86}$\BESIIIorcid{0009-0008-3419-0849},
Y.~Bai$^{62}$\BESIIIorcid{0000-0001-6593-5665}
% ...
\\
\vspace{0.2cm}
(BESIII Collaboration)\\
\vspace{0.2cm} {\it
$^{1}$ Institute of High Energy Physics, Beijing 100049, People's Republic of China\\
% ...
$^{4}$ Budker Institute of Nuclear Physics SB RAS (BINP), Novosibirsk 630090, Russia\\
% ...
$^{31}$ INFN Sezione di Ferrara, (A)INFN Sezione di Ferrara, I-44122, Ferrara, Italy; (B)University of Ferrara, I-44122, Ferrara, Italy\\
% ...
$^{39}$ Johannes Gutenberg University of Mainz, Johann-Joachim-Becher-Weg 45, D-55099 Mainz, Germany\\
% ...
$^{62}$ Southeast University, Nanjing 211100, People's Republic of China\\
% ...
$^{64}$ State Key Laboratory of Particle Detection and Electronics, Beijing 100049, Hefei 230026, People's Republic of China\\
% ...
$^{77}$ University of Science and Technology of China, Hefei 230026, People's Republic of China\\
% ...
$^{80}$ University of Turin and INFN, (A)University of Turin, I-10125, Turin, Italy; (B)University of Eastern Piedmont, I-15121, Alessandria, Italy; (C)INFN, I-10125, Turin, Italy\\
$^{81}$ Uppsala University, Box 516, SE-75120 Uppsala, Sweden\\
% ...
$^{86}$ Zhengzhou University, Zhengzhou 450001, People's Republic of China\\
\vspace{0.2cm}
$^{\dagger}$ Deceased\\
% ...
$^{c}$ Also at the Novosibirsk State University, Novosibirsk, 630090, Russia\\
% ...
}
}

\begin{abstract}
Based on $(2.712\pm0.014)\times10^9$ $\psi(3686)$ events collected with the BESIII detector, the decays $\chi_{cJ} \to p \bar{p} K^{0}_{S} K^{0}_{S}$ ($J=0,1,2$) are observed for the first time with statistical significances exceeding $5\sigma$.
The measured branching fractions are
	$\mathcal{B}(\chi_{c0}\to p \bar p K^{0}_{S} K^{0}_{S})=(6.94\pm0.30\pm0.38)\times10^{-5}$,
	$\mathcal{B}(\chi_{c1}\to p \bar p K^{0}_{S} K^{0}_{S})=(1.30\pm0.12\pm0.08)\times10^{-5}$,
	and $\mathcal{B}(\chi_{c2}\to p \bar p K^{0}_{S} K^{0}_{S})=(2.54\pm0.17\pm0.15)\times10^{-5}$,
where the first uncertainties are statistical and the second systematic.
In addition, no significant structures are observed in the $p\bar p$, $p K^0_S$, and $\bar p K^0_S$ mass spectra.
\end{abstract}

\maketitle

\section{Introduction}
Precise measurements of charmonium decay properties are important for testing perturbative quantum chromodynamics (QCD) predictions and constraining non-perturbative QCD models~\cite{pqcd1,pqcd2,pqcd3}.
The $P$-wave charmonium states $\chi_{cJ}$ ($J=0,1,2$) are ideal for such studies, as they are abundantly produced via radiative transitions from $\psi(3686)$. Moreover, these states exhibit very high hadronization efficiency in their decays, with an exceptionally strong tendency to decay into light hadrons~(such as $\pi$ and $K$ mesons).
According to the Particle Data Group (PDG)~\cite{ref::pdg2022}, the total branching fraction of $\chi_{cJ}$ ($J=0,1,2$) into light hadrons is approximately 70\%, making them excellent laboratories for studying the interplay between perturbative quark-gluon dynamics and non-perturbative hadron formation.

The decays $\chi_{cJ} \to p \bar{p} K^+ K^-$ have been observed with branching fractions ($\mathcal{B}$) at $10^{-4}$ level~\cite{bes2-ppbarkk}.
These imply that the $\chi_{cJ} \to p \bar{p} K_S^0 K_S^0$ decays are hopefully detectable at BESIII.
In addition, an enhancement  structure was observed near the $p\bar{p}$ invariant-mass threshold in the radiative decay of $J/\psi \to \gamma p \bar{p}$~\cite{J2003,M2012}.
Various theoretical models have been proposed to explain this enhancement, but no definitive conclusion has been reached~\cite{lilun1,lilun2,lilun3}.
Studies of the decays of $\chi_{cJ}\to p \bar p K^0_SK^0_S$ present additional information of possible $p\bar p$ threshold structures,
and also offer an opportunity to search for excited baryon states in the $pK^0_S$ or $\bar p K^0_S$ systems.

The BES experiment performed the first search for the $\chi_{c0,1,2} \to p \bar{p} K_S^0 K_S^0$ decays in 2006~\cite{bes2-ppkk},
yielding upper limits on their branching fractions at 90\% confidence level, corresponding
to $8.8 \times10^{-4}$, $4.5\times10^{-4}$ and $7.9\times10^{-4}$, respectively.
In this paper, we report the first observations and the branching fraction measurements of
$\chi_{c0,1,2}\to p \bar p K^{0}_{S} K^{0}_{S}$ by analyzing $(2.712\pm0.014)\times10^9$ $\psi(3686)$ events~\cite{psip-num-inc} collected with the BESIII detector.
In addition, the $p\bar p$, $p K^0_S$, and $\bar p K^0_S$ mass spectra are also examined.

\section{BESIII DETECTOR AND MONTE CARLO SIMULATION}
\label{sec:BES}
The BESIII detector~\cite{ref::detector} records symmetric $e^+e^-$ collisions provided by the BEPCII storage ring~\cite{ref::collider}
in the center-of-mass energy range from 1.84 to 4.95~GeV, with a peak luminosity of $1.1 \times 10^{33}\;\text{cm}^{-2}\text{s}^{-1}$ achieved at $\sqrt{s} = 3.773\;\text{GeV}$.

The cylindrical core of the BESIII detector covers 93\% of the full solid angle and consists of a helium-based multilayer drift chamber~(MDC),
a plastic scintillator time-of-flight system~(TOF), and a CsI(Tl) electromagnetic calorimeter~(EMC), which are all enclosed in a superconducting
solenoidal magnet providing a 1.0~T magnetic field. The solenoid is supported by an octagonal flux-return yoke with resistive plate counter muon
identification modules interleaved with steel. The charged-particle momentum resolution at $1~{\rm GeV}/c$ is $0.5\%$, and the ${\rm d}E/{\rm d}x$
resolution is $6\%$ for electrons from Bhabha scattering. The EMC measures photon energies with a resolution of $2.5\%$ ($5\%$) at $1$~GeV in the barrel
(end-cap) region. The time resolution in the TOF barrel region is 68~ps, while that in the end-cap region was 110~ps.
The end-cap TOF system was upgraded in 2015 using multi-gap resistive plate chamber technology, providing a time resolution of 60~ps~\cite{Tof1,Tof2,Tof3}.
About $83\%$ of the data used  in this analysis benefits from this upgrade.

Simulated samples produced with a {\sc
geant4}-based~\cite{Geant4} Monte Carlo (MC) package, which
includes the geometric description of the BESIII detector and the
detector response, are used to determine detection efficiencies
and to estimate backgrounds. The simulation models the beam
energy spread and initial state radiation (ISR) in the $e^+e^-$
annihilations with the generator {\sc
kkmc}~\cite{Jadach01}.

The inclusive MC sample includes the production of the
$\psi(3686)$ resonance, the ISR production of the $J/\psi$, and
the continuum processes incorporated in {\sc
kkmc}~\cite{Jadach01}.
All particle decays are modelled with {\sc
evtgen}~\cite{Lange01,Lange02} using branching fractions
either taken from the PDG~\cite{ref::pdg2022}, when available,
or otherwise estimated with {\sc lundcharm}~\cite{Lundcharm00,Lundcharm01}.
Final state radiation
from charged final state particles is incorporated using the {\sc
photos} package~\cite{PHOTOS}.
The signal MC events of the $\psi(3686) \to \gamma \chi_{cJ} (J=0,1,2)$ decays follow the angular distribution of $1 + \lambda \rm cos^2\theta_{\gamma} $, where $\theta_{\gamma}$
denotes the polar angle of the radiative photon, and $\lambda$
takes the values 1, $-1/3$, and $1/13$ for $J =$ 0, 1, and
2, respectively~\cite{p2g}.
The other decays are generated by the phase space (PHSP)
model.

\section{EVENT SELECTION}
\label{sec:selection}

We reconstruct events from the charmonium transitions $\psi(3686) \to \gamma \chi_{cJ}$ followed by the hadronic decays $\chi_{cJ} \to p \bar{p} K^0_S K^0_S$. Signal events are required to contain at least six charged tracks and at least one photon candidate.

Photon candidates are identified using showers in the EMC. The deposited energy of each shower must be more than 25~MeV in the barrel region ($|\cos \theta|< 0.80$, where $\theta$
denotes the polar angle with respect to the $z$ axis, the symmetry axis of the MDC) more than 50~MeV in the end-cap region ($0.86 <|\cos \theta|< 0.92$).
To exclude showers that originate from charged tracks, the angle subtended by the EMC shower and the position of the closest charged track at the EMC must be greater than 10 degrees as measured from the interaction point (IP).
In addition, the time information from the shower in the calorimeter relative to the event start time has to be less than 700 ns.
For charged tracks not originating from $K^0_S$ decays, the distance of closest approach to the IP must be less than 10~cm along the $z$-axis, $|V_{z}|$, and less than 1cm in the transverse plane, $|V_{xy}|$.
Particle identification (PID) for charged tracks combines d$E$/d$x$ and TOF measurements to calculate likelihood values $\mathcal{L}(h)$ for each hadron hypothesis ($h$=$p$, $\pi$, and $K$).

Each $ K^0_S$ candidate is reconstructed from two oppositely charged tracks satisfying $|V_{z}|<$20 cm.
The two charged tracks are assigned as $\pi^+\pi^-$ without imposing further PID criteria. They are constrained via vertex fit and second vertex fit to originate
from a common vertex and are required to have an invariant mass within $|M_{\pi^+ \pi^-}-m_{K^{0}_{S}}|<0.012~\mathrm{GeV/}$${c}^2$, which corresponds to $\pm 3\sigma$ where $m_{K^{0}_{S}}$ is the $K^0_S$ nominal mass~\cite{ref::pdg2022}.
The decay length of the  $ K^0_S$  candidate is required to be greater than twice the vertex resolution away from the IP.

To suppress high background containing $\Lambda$ or $\bar \Lambda$ in the final state, the $p\pi^-$ and $\bar p \pi^+$ invariant masses are required
to be outside the region of $\pm$ 0.012 GeV/$c^{2}$ around the $\Lambda$ nominal mass~\cite{ref::pdg2022}.

A four-momentum conservation constraint (4C) kinematic fit is applied to the events under the hypothesis of $e^+ e^- \to \gamma p \bar {p} K_S^0 K_S^0$, where the information on the two $K^0_S$ candidates is derived from the above second vertex fit. If multiple combinations survive in an event, the one with the smallest $\chi_{\rm 4C}^{2}$ is retained. The $\chi^2_{\rm 4C}$ selection is optimized using the figure of merit (FOM) $\mathit{S}/\sqrt{\mathit{S}+\mathit{B}}$, where $S$ is the normalized signal yield from MC simulation (scaled by the measured branching fractions) and $B$ is the normalized background from the inclusive MC sample (scaled to data). From this optimization, candidate events are required to satisfy $\chi^2_{\rm 4C} < 50$.

\begin{figure}[htbp]
	\centering
	\includegraphics[width=8.5cm]{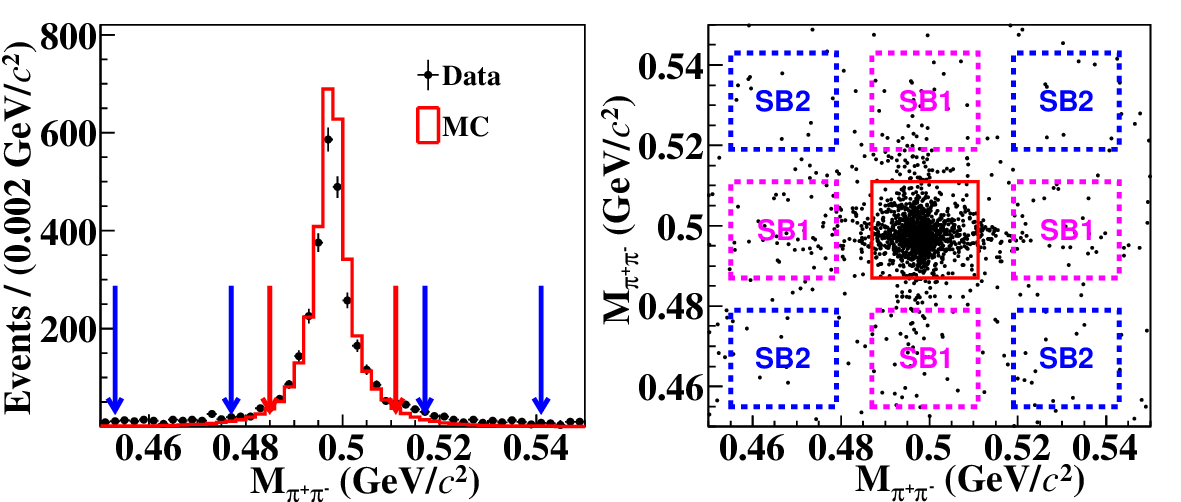}
	\caption {(left)~The $M_{\pi^+\pi^-}$ distribution of the $\chi_{cJ}\to p\bar p K^0_SK^0_S$ candidates in
data, where both $K_S^0$ candidates are included. The red (blue) solid arrows indicate the 1D $K^0_S$ signal (sideband) regions. (Right) The distribution of $M_{\pi^+\pi^-}$ versus $M_{\pi^+\pi^-}$ (with the two $\pi^+\pi^-$ combinations randomly assigned) for $\chi_{cJ} \to p\bar{p} K^0_S K^0_S$ candidates in data. The red solid box denotes the 2D signal region, while the pink and blue dot-dashed boxes indicate the 2D SB1 and SB2 regions, respectively.}
	\label{tab:datakso}
\end{figure}

Non-$K_S^0$ $\pi^+\pi^-$ combinations in the $K^0_S$ signal region may form peaking backgrounds around $M_{\chi_{cJ}}$ in the $M_{p\bar p K^0_SK^0_S}$ distributions.
This kind of peaking background contribution is estimated with events in the one-dimensional~(1D) $K^0_S$ sideband region of
$(0.454,0.478)\cup(0.518,0.542)~{\rm GeV}/c^2$.
Two-dimensional (2D) $K^0_SK^0_S$ signal and sideband regions~(SB) are also defined: the signal region contains both $\pi^+\pi^-$ pairs within the 1D $K^0_S$ signal region; SB1 contains one $\pi^+\pi^-$ pair in the 1D $K^0_S$ sideband region and one in the 1D $K^0_S$ signal region; and SB2 contains both pairs in the 1D $K^0_S$ sideband regions. Figure~\ref{tab:datakso} shows the 1D and 2D $\pi^+\pi^-$ invariant-mass distributions along with the $K^0_S$ signal and sideband regions.

Other potential background components from $\psi(3686)$ decays, which can not be described by the sidebands, are studied by analyzing the inclusive MC sample with the generic event type analysis tool, TopoAna~\cite{topo}. No significant peaking background reactions are observed.
Additionally, we examine data taken at $\sqrt{s} = 3.650$~GeV, corresponding to an integrated luminosity of 445.5~pb$^{-1}$~\cite{lum}, to estimate potential continuum production background. It is found to be negligible.

The net signal yields for the $\chi_{cJ}\to p\bar p K^0_SK^0_S$ decays are determined as
\begin{equation*}
N_{\mathrm{net}}^{\chi_{cJ}}=N_{\mathrm{sig}}-\frac{1}{2}N_{\mathrm{SB1}}+\frac{1}{4}N_{\mathrm{SB2}},
\end{equation*}	
where $N_{\mathrm{sig}}$, $N_{\mathrm{SB1}}$, and $N_{\mathrm{SB2}}$ are the numbers of events in the 2D signal, SB1, and SB2 regions, respectively.
To determine $N_{\mathrm{sig}}$, $N_{\mathrm{SB1}}$, and $N_{\mathrm{SB2}}$,
unbinned maximum likelihood fits are performed on the corresponding $M_{p \bar p K^0_S K^0_S}$ distributions.

\begin{figure*}[htbp]
	\centering
	\includegraphics[width=0.32\textwidth]{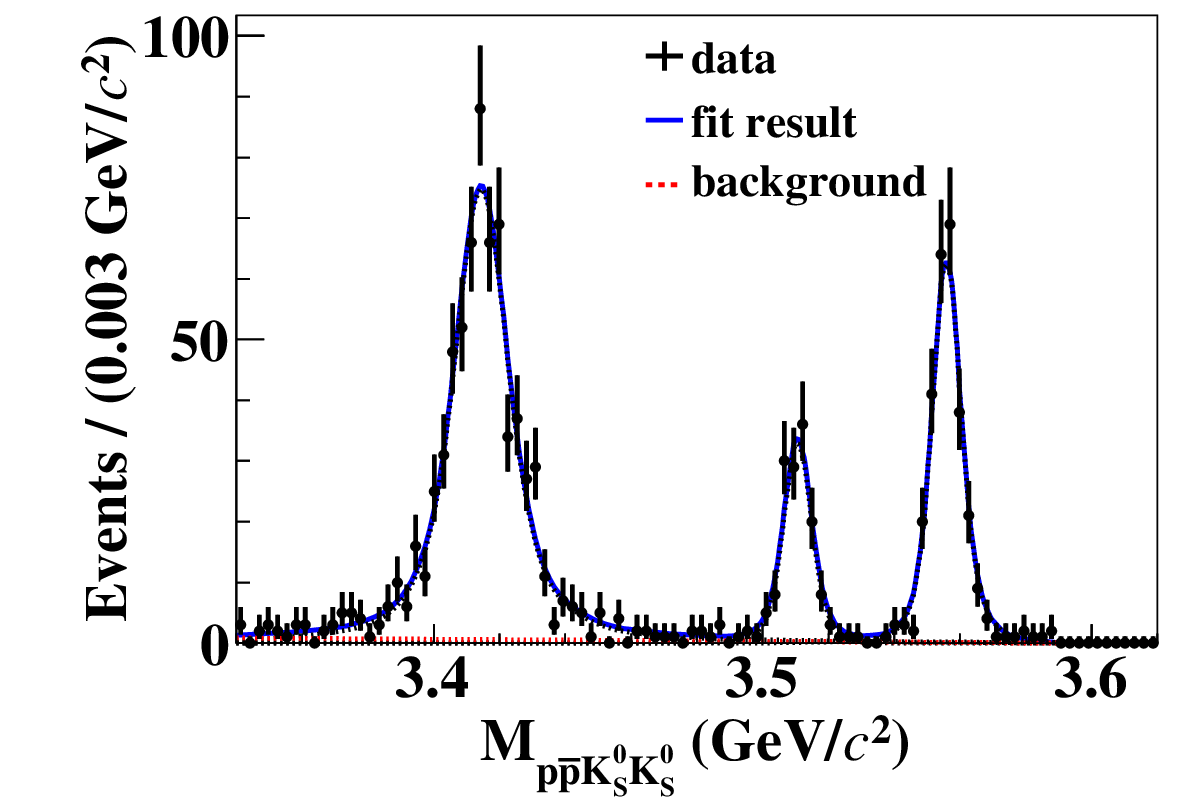}
	\includegraphics[width=0.32\textwidth]{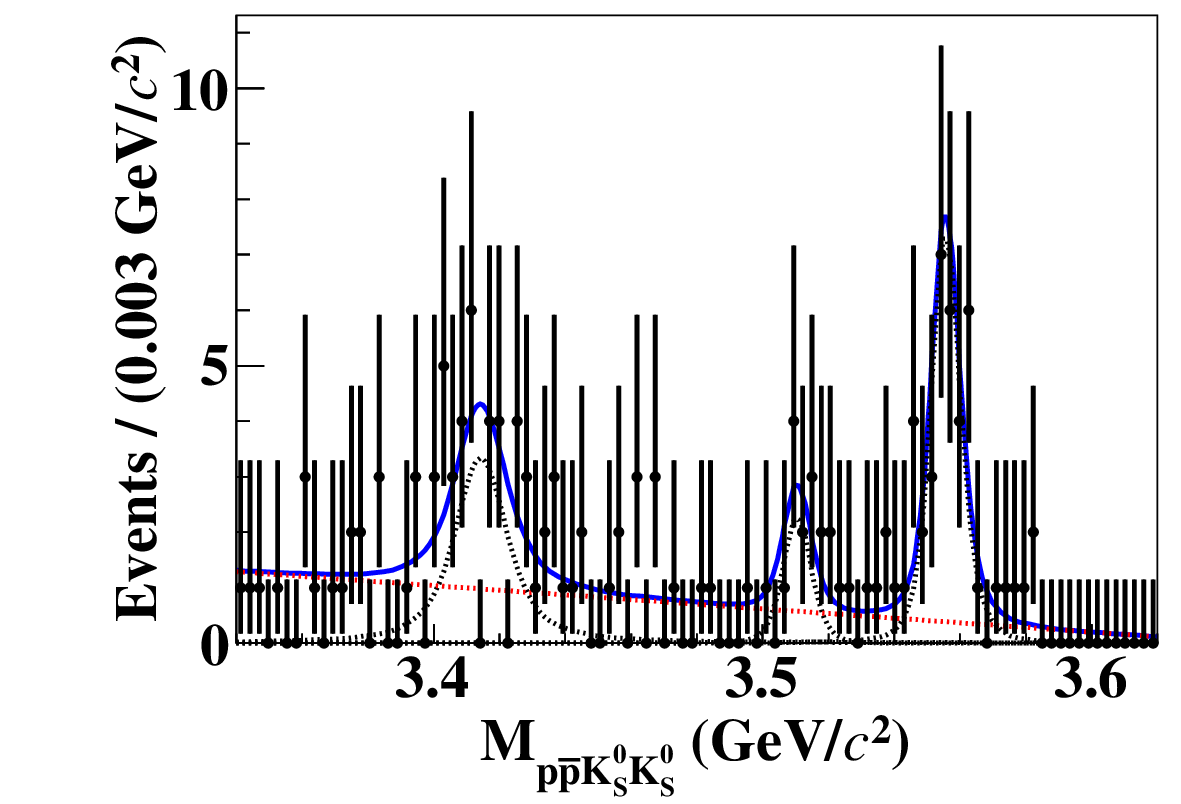}
	\includegraphics[width=0.32\textwidth]{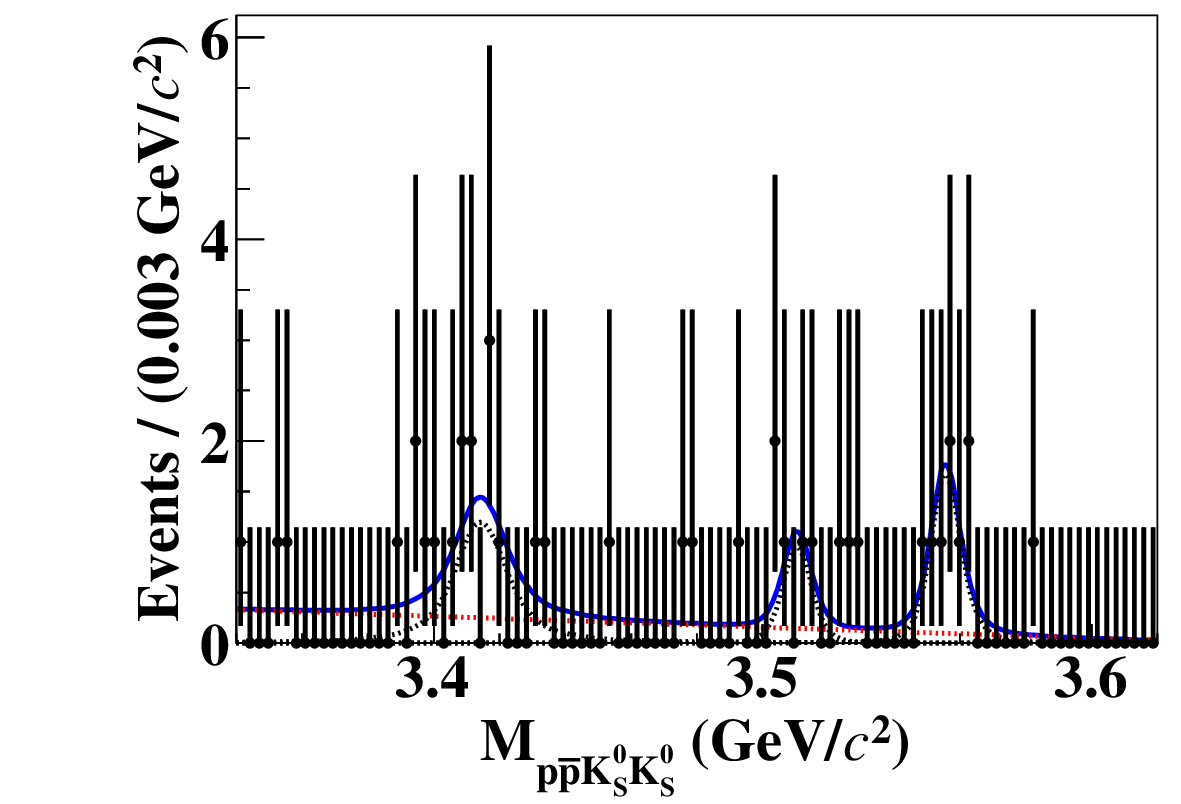}
	\caption{Fits to the $M_{p \bar p K^{0}_{S} K^{0}_{S} }$ distributions of the accepted candidates for $\chi_{cJ}\to p \bar p K^{0}_{S} K^{0}_{S} $
		in data in the $ K^{0}_{S}$ signal region~(left), SB1 region~(middle) and SB2 region~(right).
		The dots with error bars are data, the solid curves are the fit results, the dashed curves are the fitted combinatorial background shapes,
	    and the black dash lines are the signal.}
	\label{tab:sigfit}
\end{figure*}

The signal yields are obtained by performing unbinned maximum likelihood fits to the mass spectra
of the selected $M_{p\bar p K_S^0 K_S^0}$ in the 2D $K_S^0$ signal region or 2D sideband regions.
In the fit, the signal shapes of $\chi_{cJ}$ are modeled using the MC simulation
with a Gaussian resolution function, with free parameters, to account for differences in resolution between data and MC.
The background is described by a linear function.
The signal shapes in the $K^0_SK^0_S$ signal and sideband regions share the same Gaussian function.
Figure \ref{tab:sigfit} shows the fit results. Based on these fits, the net signal yields ($N^{\rm net}_{\chi_{cJ}}$) of $\chi_{c0}$, $\chi_{c1}$ and $\chi_{c2}$ are determined to be $680.0\pm29.2$, $133.4\pm12.2$ and $266.9\pm17.5$, respectively.
The statistical significance of each signal decay is found to be much higher than $5\sigma$, which is evaluated by comparing the likelihoods with and without the signal component and taking into account the change in the number of degrees of freedom.

\begin{figure*}
	\centering
	\scalebox{1.035}{
	\hspace{-0.25cm}
	\includegraphics[width=0.24\textwidth]{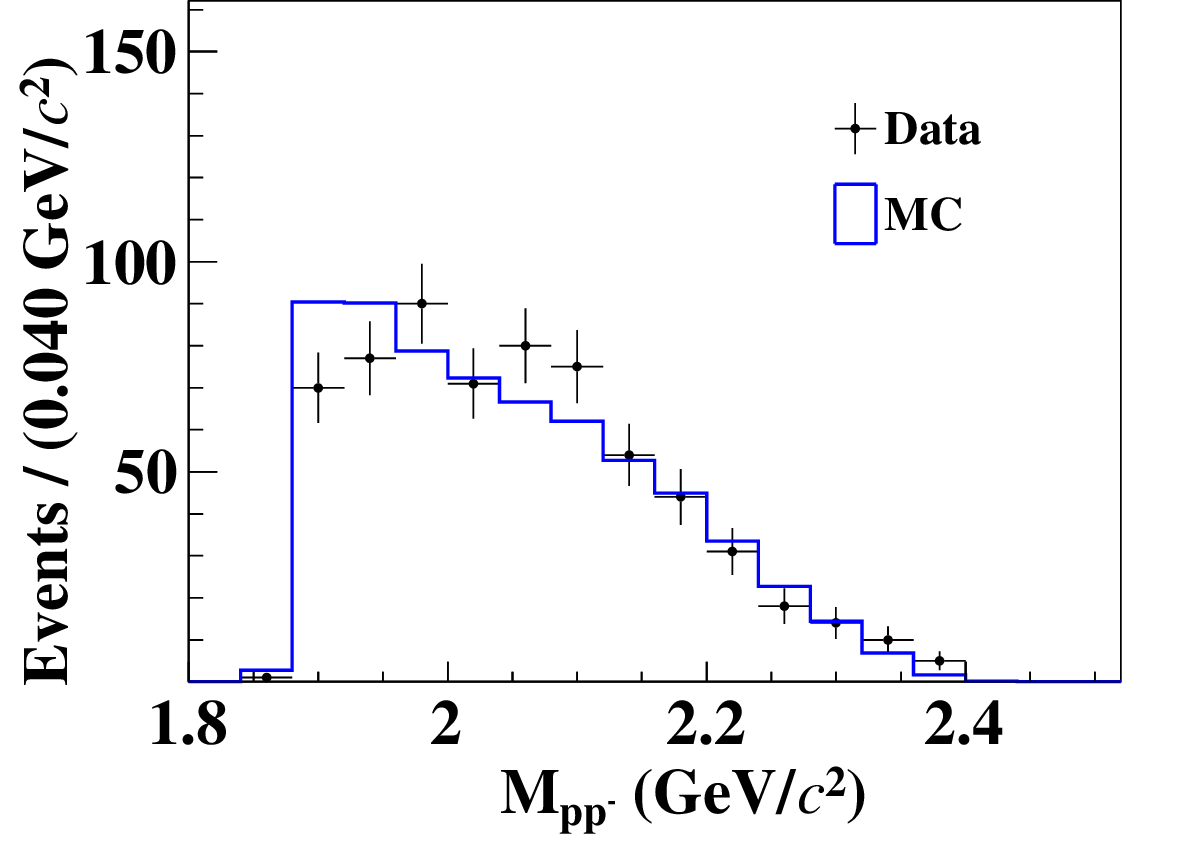}
	\includegraphics[width=0.24\textwidth]{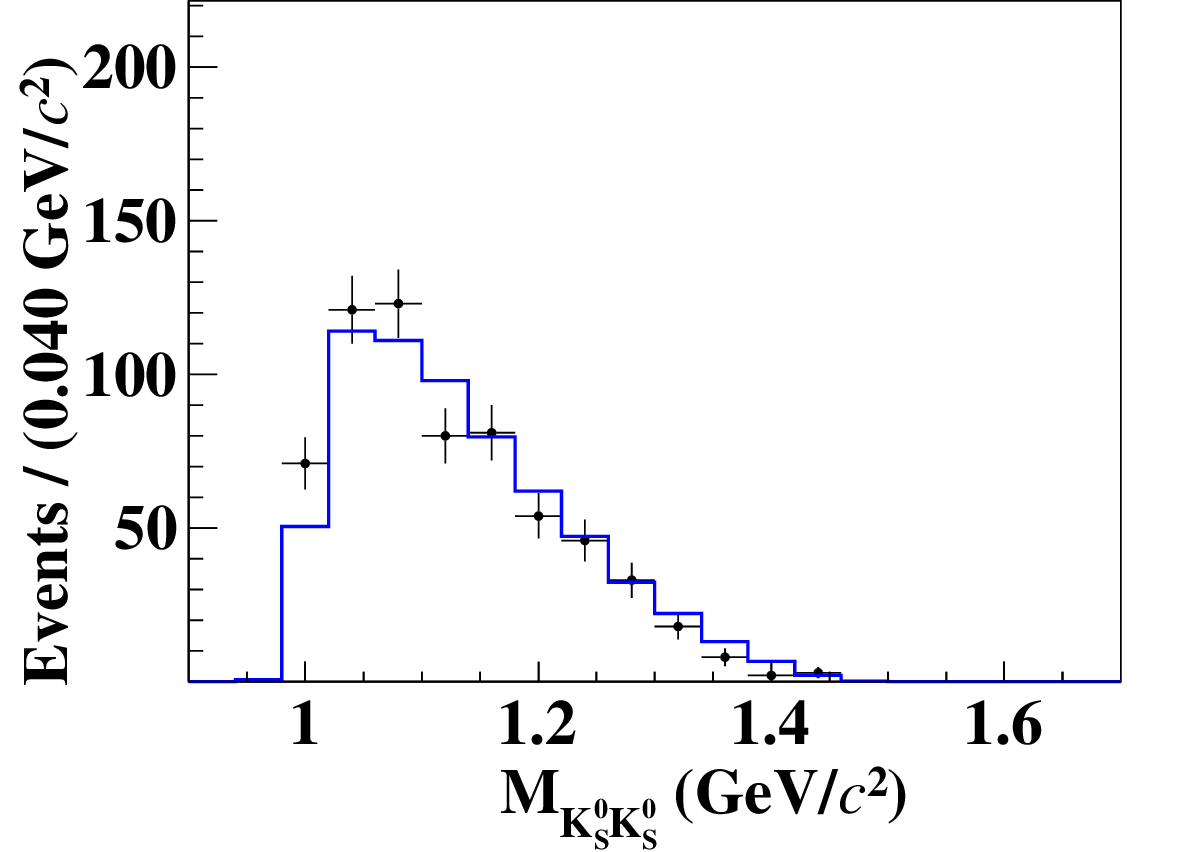}
	\includegraphics[width=0.24\textwidth]{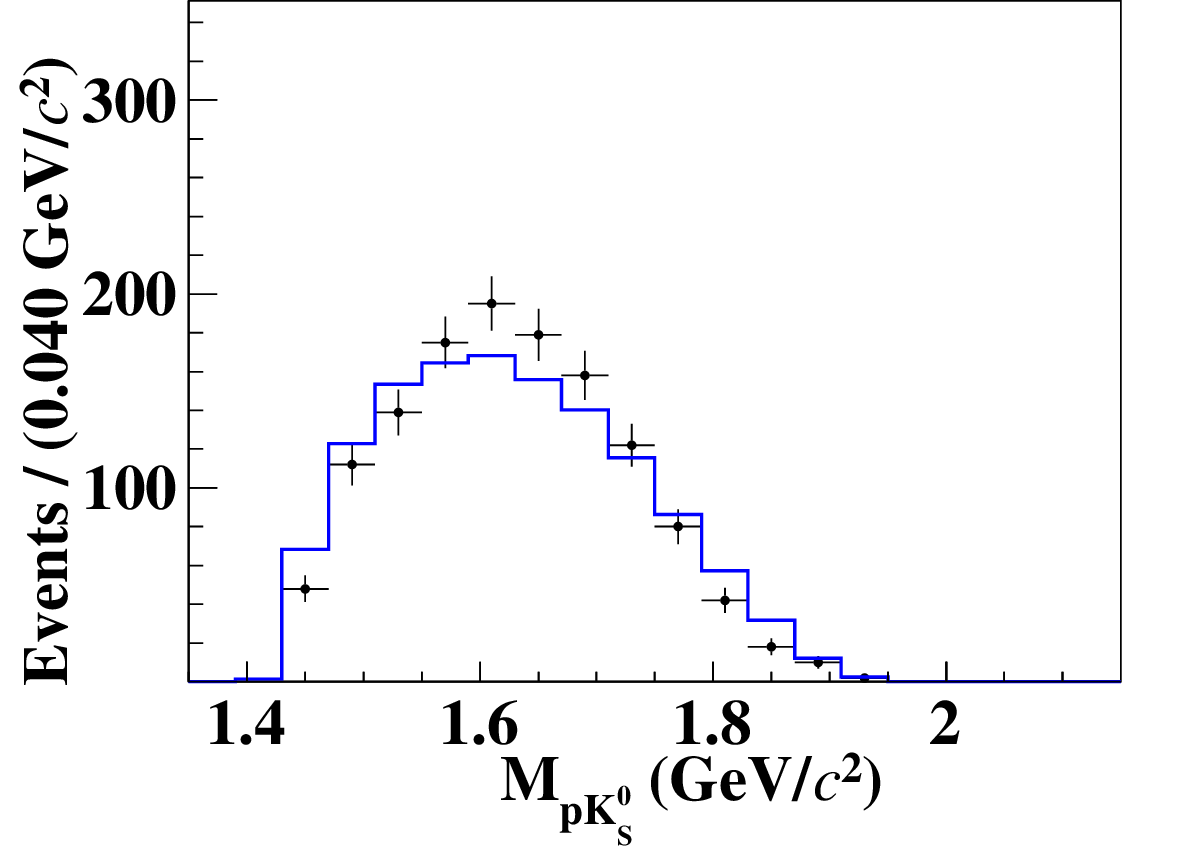}
	\includegraphics[width=0.24\textwidth]{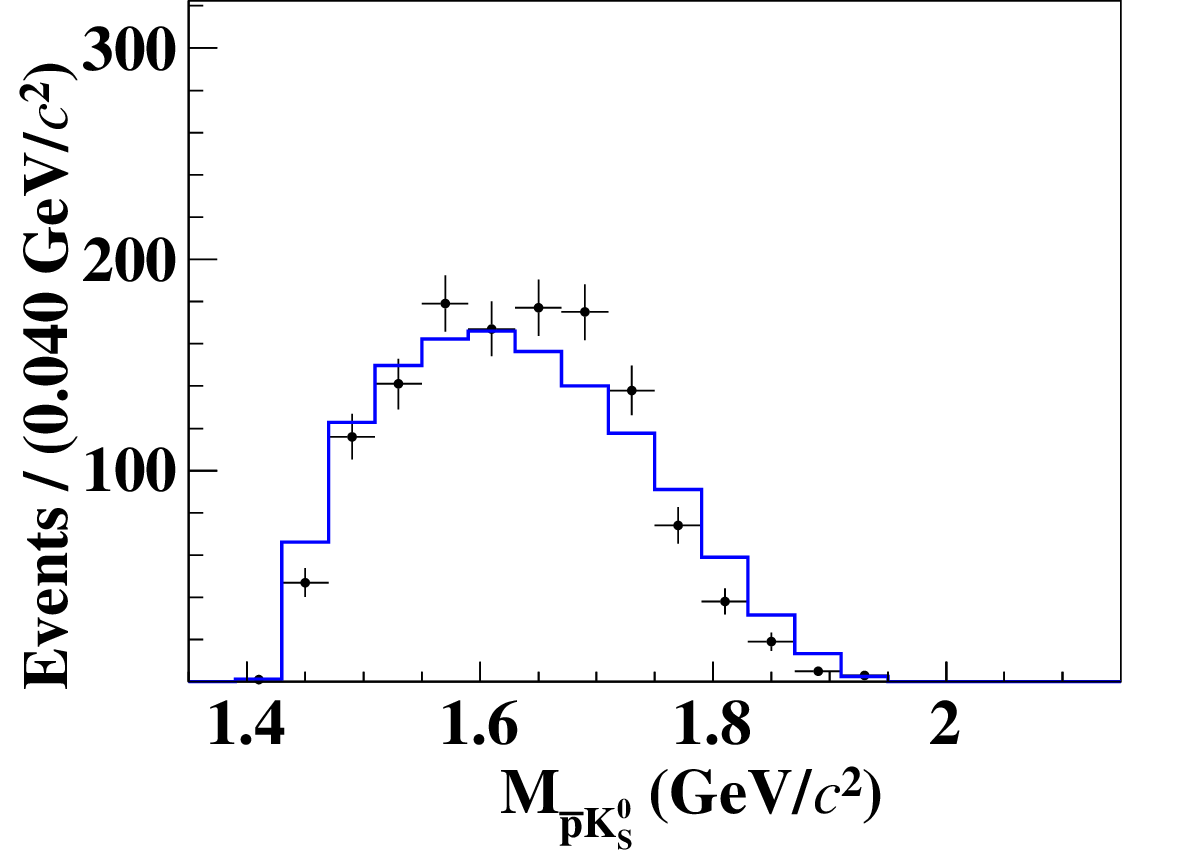}
	}
\\
\vspace{-0.67cm}
\scalebox{1.035}{
\hspace{-0.25cm}
	\includegraphics[width=0.24\textwidth]{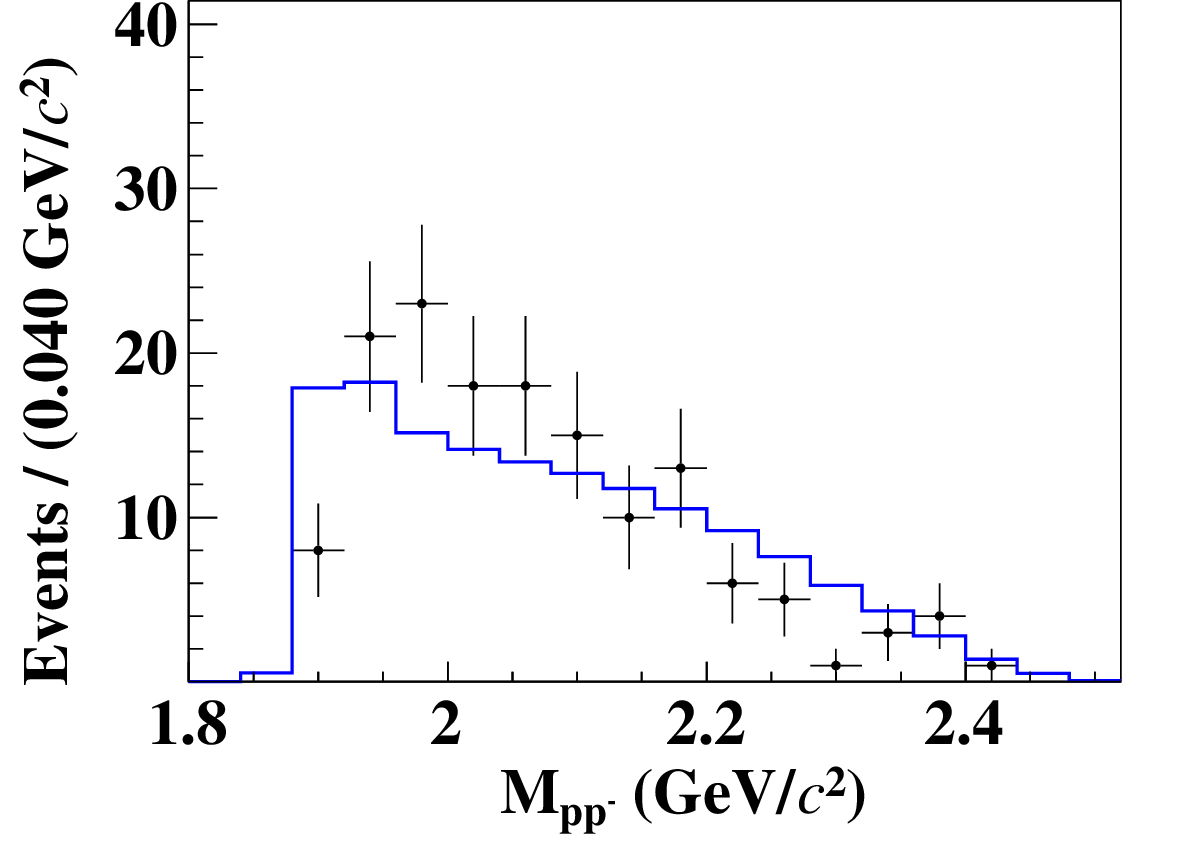}
	\includegraphics[width=0.24\textwidth]{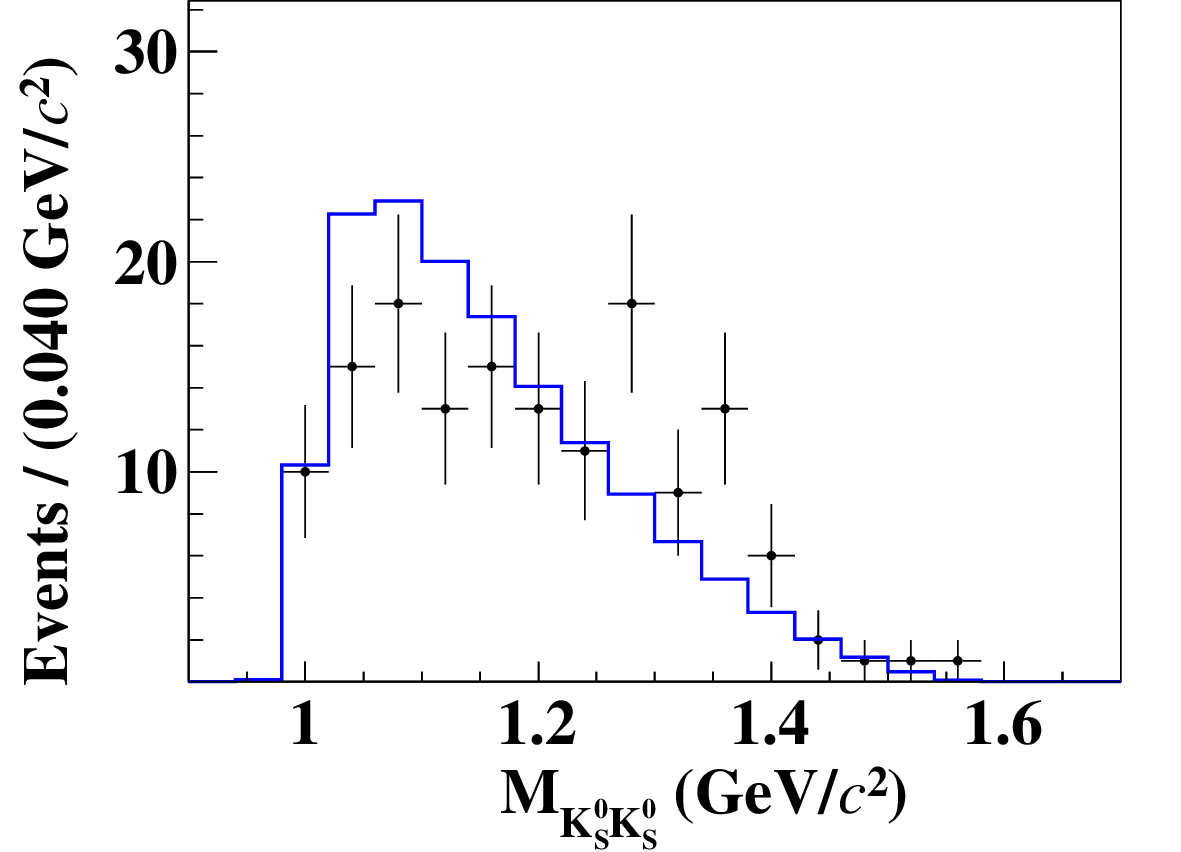}
	\includegraphics[width=0.24\textwidth]{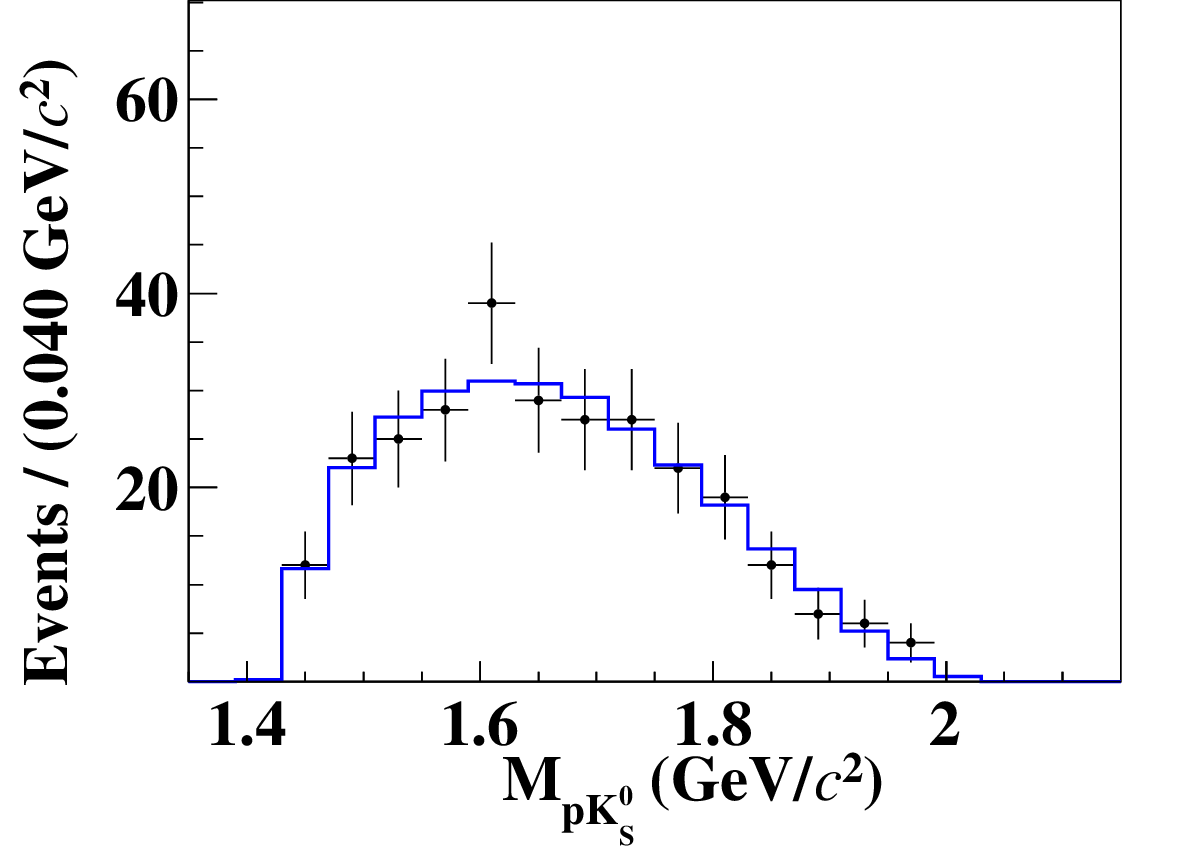}
	\includegraphics[width=0.24\textwidth]{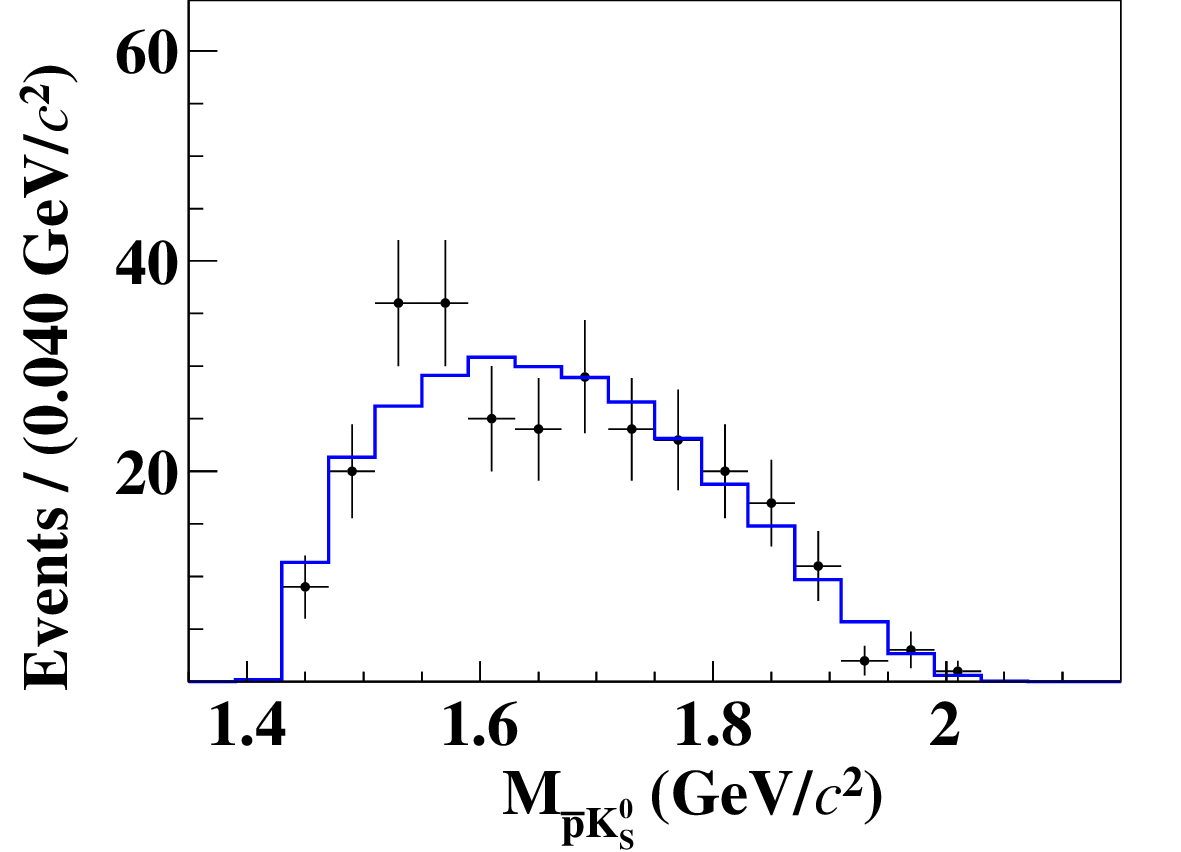}
}
\\
\vspace{-0.67cm}
\scalebox{1.035}{
\hspace{-0.25cm}
	\includegraphics[width=0.24\textwidth]{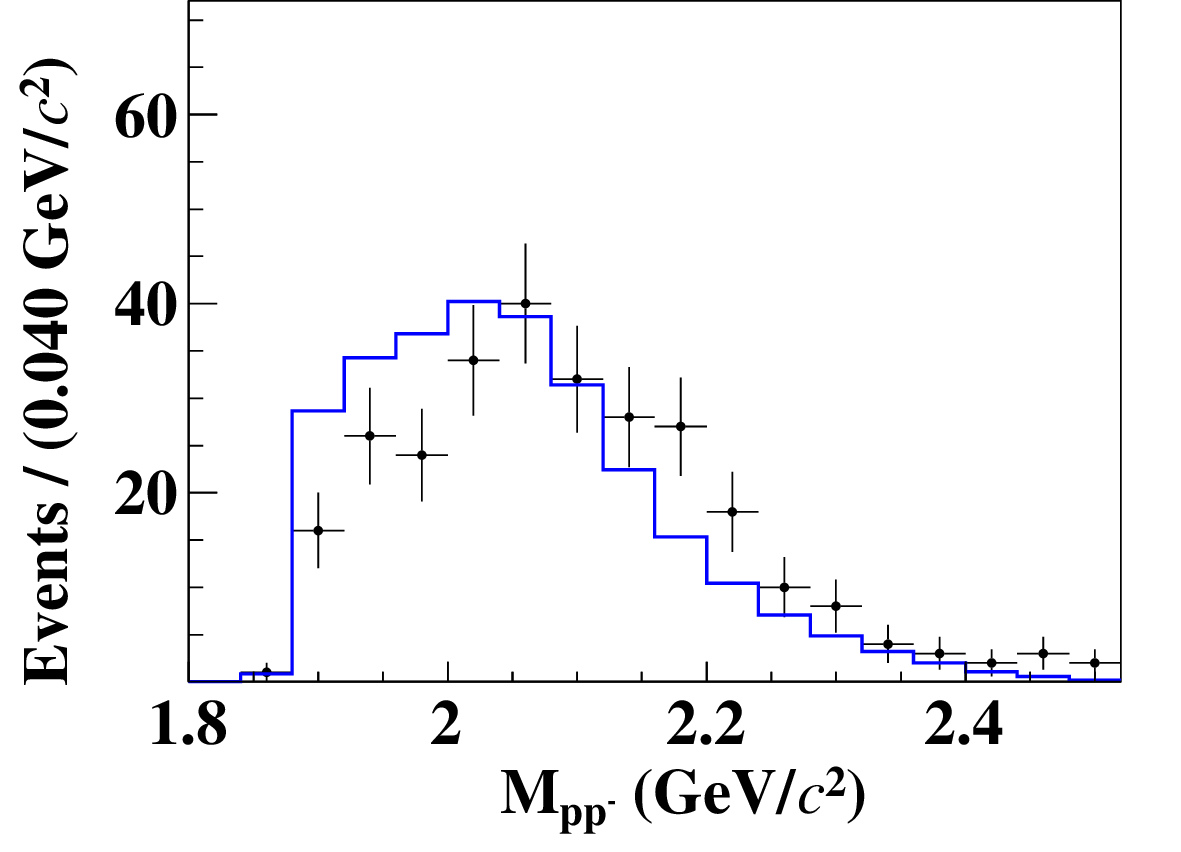}
    \includegraphics[width=0.24\textwidth]{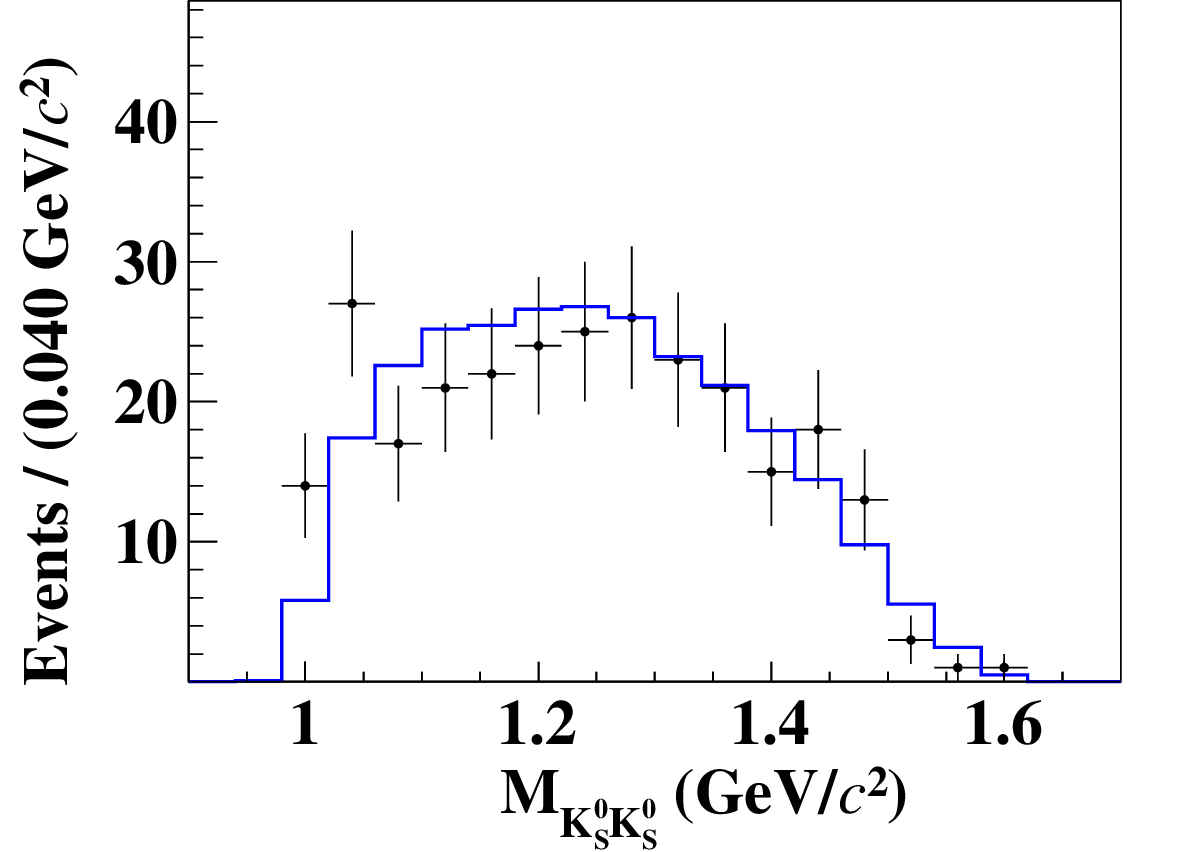}
    \includegraphics[width=0.24\textwidth]{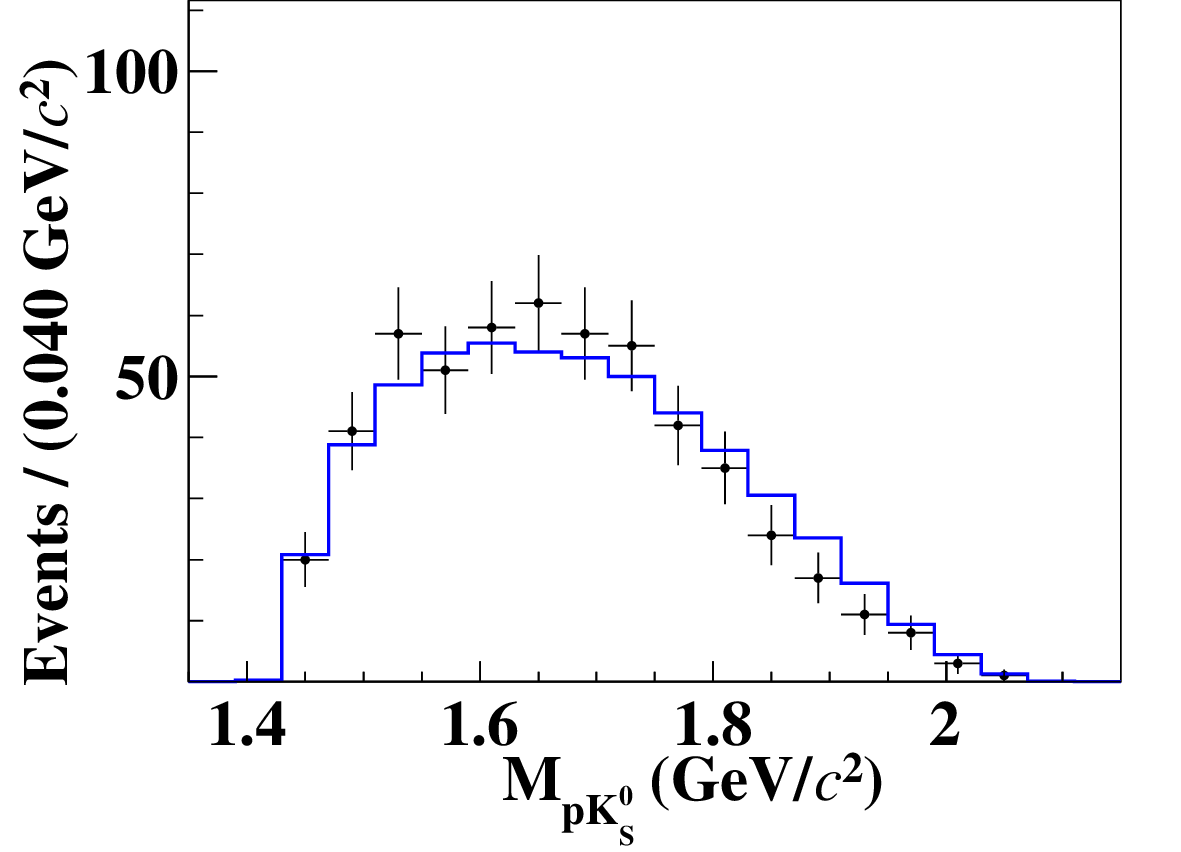}
    \includegraphics[width=0.24\textwidth]{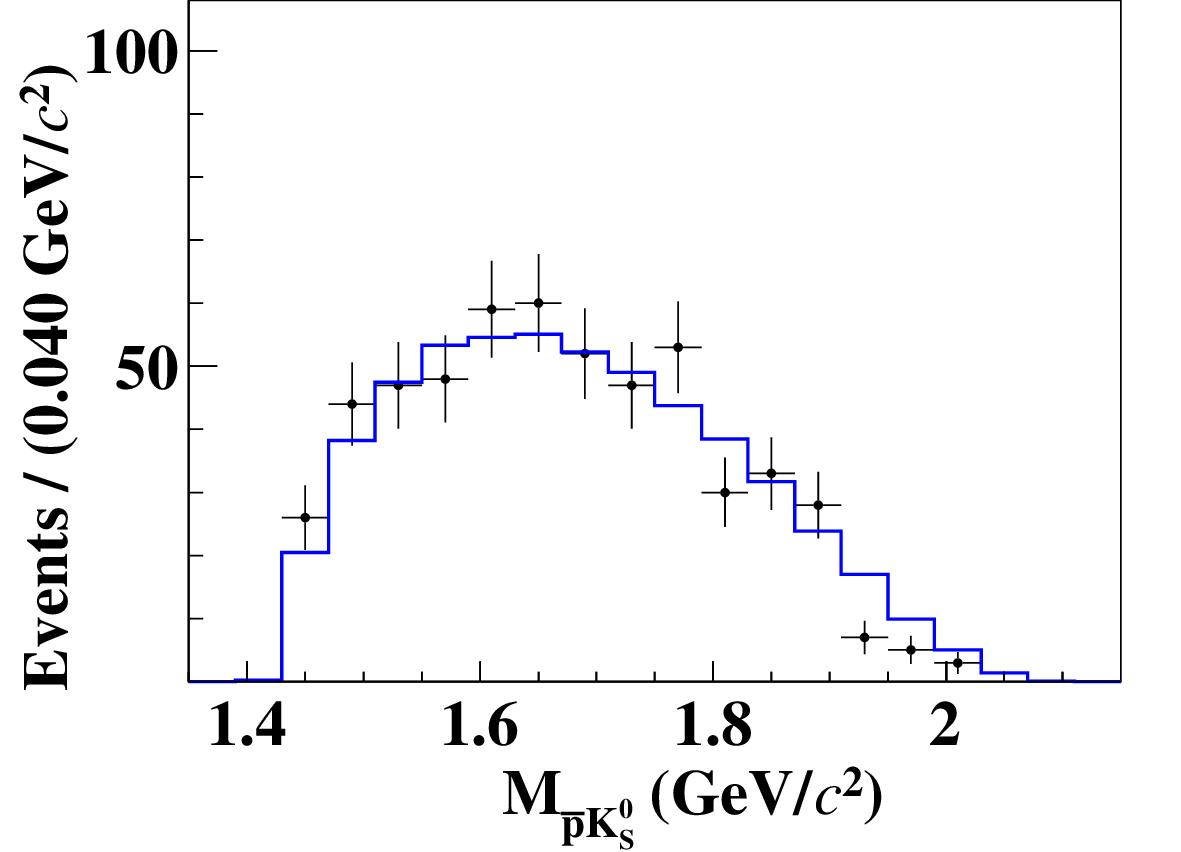}
}
	\caption{
		Comparisons of the distributions the invariant masses of two-body combinations
		of the candidates for (top) $\chi_{c0}$, (middle) $\chi_{c1}$, and (bottom) $\chi_{c2}$ decaying into  $p\bar p K^0_S K^0_S$.
		The black dots with error bars are data and the blue histograms are the signal MC sample. Events have been required to fall in
		the $\chi_{c0}$, $\chi_{c1}$, and $\chi_{c2}$ signal regions defined as $[3.385, 3.445]$~GeV/$c^2$, $[3.490, 3.530]$~GeV/$c^2$
		and  $[3.535, 3.575]$~GeV/$c^2$, respectively. There are two entries for $M_{p K^0_S}$  and $M_{\bar p K^0_S}$.
		}
	\label{tab:con0}
\end{figure*}

Figure \ref{tab:con0} presents the invariant mass distributions of two-body combinations accepted $\chi_{c0,1,2}\to p\bar p K^0_SK^0_S$ candidates.
Good consistencies between data and MC simulations can be seen, demonstrating the reliability of the signal MC samples.
No obvious structures in the $p\bar p$, $pK^0_S$ and $\bar p K^0_S$ invariant mass spectra have been observed.

\section{Branching fraction}

The product branching fractions of $\psi(3686) \to \gamma \chi_{cJ}$ and $\chi_{cJ} \to p \bar{p} K^0_S K^0_S$ are calculated as
\begin{equation}
\begin{split}
&\mathcal B(\psi(3686) \to \gamma \chi_{cJ}) \cdot \mathcal B(\chi_{cJ} \to p \bar{p} K_S^{0} K_S^{0}) \\
&\quad = \frac{N^{\rm net}_{\chi_{cJ}}}{N_{\psi(3686)} \cdot\epsilon\cdot\mathcal B( K^0_{S} \to \pi^+ \pi^-)^2 },
\end{split}
\end{equation}
where $N_{\psi(3686)}$ is the total number of $\psi(3686)$ events in data, $\epsilon$ is the detection efficiency and  $\mathcal B(K^0_{S} \to \pi^+ \pi^-)$ is
the branching fraction of $K_S^0 \to \pi^+\pi^-$ quoted from the PDG~\cite{ref::pdg2022}.

According to Eq. (1),
we obtain the product branching fractions, ${\mathcal B}(\psi(3686)\to \gamma\chi_{cJ}) \cdot \mathcal{B}(\chi_{cJ} \to p \bar{p} K_S^{0} K_S^{0})$, and by
combining our results with the world average values of ${\mathcal B}(\psi(3686)\to \gamma\chi_{cJ})$~\cite{Lundcharm00}, we obtain $\mathcal{B}(\chi_{cJ} \to p \bar{p} K_S^{0} K_S^{0})$.
All relevant quantities are summarized in Table~\ref{tab:branching}.
\begin{table*}
	\centering
	\caption{The quantities used for determining the branching fractions of $\chi_{cJ}\to p \bar p K^{0}_{S} K^{0}_{S}$,
	where the first uncertainties are statistical and the second systematic.}
	\small
	\begin{tabular}{cccc}
		\hline
		\hline
		& $\chi_{c 0}$ & $\chi_{c 1}$ & $\chi_{c 2}$ \\ \hline
		$N_{\text{sig}}$  & $692.1\pm 28.9$ & $137.1 \pm 12.1$ & $281.5\pm 17.2$ \\
		$N_{\text{SB1}}$ & $29.5\pm 8.3$    &  $9.4\pm3.9$   & $32.9\pm6.6$ \\
		$N_{\text{SB2}}$ &  $10.6\pm4.5$    & $4.0\pm2.5$   &  $7.5\pm3.0$ \\
		$N^{\rm net}_{\chi_{cJ}}$  &  $680.0\pm29.2$ &  $133.4\pm12.2$ &  $266.9\pm17.5$ \\
		$\epsilon$~(\%)   & $7.72\pm0.03$       & $8.13\pm0.03$       & $8.67\pm0.03$ \\
		$\mathcal B(\psi(3686) \to \gamma \chi_{cJ}) \cdot \mathcal B(\chi_{cJ} \to p \bar{p} K_S^{0} K_S^{0})$
		($10^{-6}$) &$6.79\pm0.29\pm0.33$ & $1.27\pm0.12\pm0.07$ &$2.37\pm0.16\pm0.12$ \\
		$\mathcal B(\chi_{cJ}\to p\bar p K^0_SK^0_S)$ ($10^{-5}$) & $6.94 \pm 0.30\pm0.38$ & $1.30 \pm 0.12\pm0.08 $
		 & $2.54 \pm 0.17\pm0.15$ \\ \hline \hline
	\end{tabular}
	\label{tab:branching}
\end{table*}

Our analysis procedure has been validated via input-output checks with 50 simulated samples.
Each simulated sample has the same size as the real data. The same analysis procedure as the data analysis is performed on each
simulated sample. The average of the measured branching fractions agrees well with the input branching fraction,
thereby ensuring that there is no bias in the analysis procedure.

\section{SYSTEMATIC UNCERTAINTY}
\label{sec:systematics}
The systematic uncertainties in the branching fraction measurements arise from several sources, as summarized in Table~\ref{tab:systematics}, and are discussed below.
\begin{table*}[]
	\centering
  	\caption{Relative systematic uncertainties in the measurements of the branching fractions of $\chi_{cJ}\to p \bar p K^{0}_{S} K^{0}_{S}$ (in percent).}
  \begin{tabular}{lccc}
  	\hline
  	\hline
  	Source & $\chi_{c0}$ & $\chi_{c1}$ & $\chi_{c2}$  \\ \hline
  	$N_{\psi(3686)}$ & $ 0.5$ & $ 0.5$ & $ 0.5$ \\
  	Tracking & $ 2.0$ & $ 2.0$ & $ 2.0$ \\
  	PID      & $ 2.0$ & $ 2.0$ & $ 2.0$ \\
  	$K_{S}^{0}$ reconstruction & $ 3.0$ & $ 3.0$ & $ 3.0$\\
  	Photon detection & $ 1.0$ & $ 1.0$ & $ 1.0$\\
  	4C kinematic fit & $ 0.1$ & $ 0.1$ & $ 0.2$ \\
	MC model & $ 0.6$ & $ 0.4$ & $ 0.1$ \\
  	Fit to $M_{p\bar p K^0_SK^0_S}$ & $ 2.3$ & $ 3.0$ & $ 3.0$ \\
	 MC statistics & $ 0.4$ & $ 0.4$ & $ 0.4$\\
  	$\mathcal B(K_{S}^{0} \to \pi^{+} \pi^{-})$ & $ 0.2$ & $ 0.2$ & $ 0.2$ \\
	$\mathcal B(\psi(3686) \to \gamma \chi_{cJ})$   &$2.0$ & $2.5$ &$2.1$ \\ \hline
	Total for ${\mathcal B}(\psi(3686)\to \gamma \chi_{cJ})\cdot {\mathcal B}(\chi_{cJ}\to p\bar p K^0_SK^0_S)$~ &
	$4.9$  & $5.2$ & $5.2$ \\  \hline
  	Total for ${\mathcal B}(\chi_{cJ}\to p\bar p K^0_SK^0_S)$ & $ 5.3$ & $5.8$ & $5.7$ \\
  	\hline
  	\hline
  \end{tabular}
  \label{tab:systematics}
\end{table*}

The total number of $\psi(3686)$ events in data is measured to be $N_{\psi(3686)}=(2.712\pm0.014)\times10^9$ with the inclusive
hadronic events, as described in Ref.~\cite{psip-num-inc}. The systematic uncertainty of $N_{\rm \psi(3686)}$ is 0.5\%.

The tracking efficiency of protons is studied using a control sample of $J/\psi\to p\bar p\pi^+\pi^-$. The difference in tracking efficiency between data and MC simulation is assigned as the systematic uncertainty
is $1.0\%$ per track. Using the same control sample, the systematic uncertainty due to the PID of proton is also assigned to be $1.0\%$ per track~\cite{track}.

The systematic uncertainty of the $K^0_S$ reconstruction, including tracking efficiency, $K^0_S$ mass window,
vertex fit and second vertex fit, is estimated by using control samples of $J/\psi \rightarrow K^*(892)^{\pm} \bar{K}^\mp$ and $J/\psi \rightarrow \phi K^*(892)^{\pm} \bar{K}^\mp$.
The systematic uncertainty of the $K^0_S$ reconstruction is assigned to be 1.5\%~\cite{ref::kso}.

The systematic uncertainty in the photon detection is assigned as  1.0\% per photon with the control sample of $J/\psi\to\pi^+\pi^-\pi^0$~\cite{ref::gamma-recon}.

The systematic uncertainty from the 4C kinematic fit is estimated by comparing the signal efficiencies with and without helix parameter
correction, where the correction factors are taken from the previous study in Ref.~\cite{ref::helixp}.
The differences in the signal efficiencies are assigned as the systematic uncertainties, which are 0.1 \%, 0.1 \%, and 0.2 \% for $\chi_{c0,1,2}\to p \bar p K^{0}_{S} K^{0}_{S}$, respectively.

The systematic uncertainties of the MC model are estimated to be 0.6\%, 0.4\%, and 0.1\%,
for the $\chi_{c0}$, $\chi_{c1}$ and $\chi_{c2}$ decays, which are the differences between the nominal and reweighted PHSP signal MC samples. Here, the latter ones are reweighted according to their $M_{p\bar p}$ versus $M_{K^0_SK^0_S}$ in data.

The systematic uncertainties of the fit to the $M_{p \bar p K^{0}_{S} K^{0}_{S}}$ distribution are evaluated from two aspects.
The uncertainties due to the background shape are estimated by varying between first and second-order Chebyshev polynomials.
The uncertainties of the signal shape are examined by changing the nominal signal MC (convolved with a Gaussian function) to a Breit-wigner function multiplied by $E^3_\gamma$, where $E_\gamma$ is the energy of the radiative photon~($E_\gamma=\frac{m^2_{\psi{(3686)}}-m^2}{2m_{\psi(3686)}}$).
Adding these two effects in quadrature yields the systematic uncertainties to be 2.3\%, 3.0\%, and 3.0\% for $\chi_{c0,1,2}\to p \bar p  K^{0}_{S} K^{0}_{S}$,
respectively.

The uncertainties due to limited statistics of the signal MC samples
are 0.4\%, 0.4\%, and 0.4\% for  $\chi_{c0,1,2}\to p \bar p K^{0}_{S} K^{0}_{S}$.

The branching fractions of the $\chi_{cJ}$ and $K^0_S$ decays are quoted from the PDG~\cite{ref::pdg2022}.
The corresponding uncertainties are 2.0\%, 2.5\%, 2.1\% for $\psi(3686)\to \gamma\chi_{c0,1,2}$, and 0.07\% for $K^{0}_{S} \to \pi^+ \pi^-$.

The total systematic uncertainty for each signal decay is obtained by adding all contributions in quadrature.

\section{Summary}
Using $(2.712 \pm 0.014) \times 10^9$ $\psi(3686)$ events collected with the BESIII detector, the decays $\chi_{cJ} \to p \bar{p} K^{0}_{S} K^{0}_{S}$ are observed for the first time with statistical significances exceeding $5\sigma$.
The measured branching fractions of $\chi_{cJ}\to p \bar p K^{0}_{S} K^{0}_{S}$ are summarized in Table~\ref{tab:branching}.
We have also examined the $p\bar p$, $pK^0_S$, and $\bar p K^0_S$ invariant mass spectra, and no significant structures are observed.

\section{Acknowledgement}
The BESIII Collaboration thanks the staff of BEPCII (https://cstr.cn/31109.02.BEPC) and the IHEP computing center for their strong support. This work is supported in part by National Key R\&D Program of China under Contracts Nos. 2023YFA1606000, 2023YFA1606704; National Natural Science Foundation of China (NSFC) under Contracts Nos. 11635010, 11935015, 11935016, 11935018, 12025502, 12035009, 12035013, 12061131003, 12192260, 12192261, 12192262, 12192263, 12192264, 12192265, 12221005, 12225509, 12235017, 12361141819; the Chinese Academy of Sciences (CAS) Large-Scale Scientific Facility Program; the Strategic Priority Research Program of Chinese Academy of Sciences under Contract No. XDA0480600; CAS under Contract No. YSBR-101; 100 Talents Program of CAS; The Institute of Nuclear and Particle Physics (INPAC) and Shanghai Key Laboratory for Particle Physics and Cosmology; ERC under Contract No. 758462; German Research Foundation DFG under Contract No. FOR5327; Istituto Nazionale di Fisica Nucleare, Italy; Knut and Alice Wallenberg Foundation under Contracts Nos. 2021.0174, 2021.0299; Ministry of Development of Turkey under Contract No. DPT2006K-120470; National Research Foundation of Korea under Contract No. NRF-2022R1A2C1092335; National Science and Technology fund of Mongolia; Polish National Science Centre under Contract No. 2024/53/B/ST2/00975; STFC (United Kingdom); Swedish Research Council under Contract No. 2019.04595; U. S. Department of Energy under Contract No. DE-FG02-05ER41374

\end{document}